\documentclass{PoS}
\usepackage{cite}
\usepackage[utf8]{inputenc}

\title{Hadronic uncertainties in the $B\to K^*\mu^+\mu^-$ decay}

\ShortTitle{Hadronic uncertainties in the $B\to K^*\mu^+\mu^-$ decay}

\author{\speaker{Marco Ciuchini}\\
        INFN Sezione di Roma Tre,\\
        Via della Vasca Navale 84, I-00146, Rome, Italy\\
        E-mail: \email{marco.ciuchini@roma3.infn.it}}

\author{Antonio M. Coutinho\\
        Dipartimento di Matematica e Fisica, Universit\`a di Roma Tre and INFN Sezione di Roma Tre,\\
        Via della Vasca Navale 84, I-00146, Rome, Italy\\
        E-mail: \email{antonio.coutinho@roma3.infn.it}}

\author{Marco Fedele\\
        Laboratoire de Physique Th\'eorique, B\^at.~210, \\
        Universit\'e Paris Sud, F-91405 Orsay cedex, France\\
	    E-mail: \email{marco.fedele@uniroma1.it}}

\author{Enrico Franco\\
	INFN Sezione di Roma,\\
	Piazzale Aldo Moro 2, I-00185, Rome, Italy\\
	E-mail: \email{enrico.franco@roma1.infn.it}}

\author{Ayan Paul\\
	DESY, Notkestra{\ss}e 85, D-22607 Hamburg, Germany and\\
	Institut f\"ur Physik, Humboldt-Universit\"at zu Berlin, D-12489 Berlin, Germany\\
	E-mail: \email{ayan.paul@desy.de}}

\author{Luca Silvestrini\\
	CERN, 1211 Geneva 23, Switzerland and INFN Sezione di Roma,\\
	Piazzale Aldo Moro 2, I-00185, Rome, Italy\\
	E-mail: \email{luca.silvestrini@roma1.infn.it}}

\author{Mauro Valli\\
	Department of Physics and Astronomy, University of California, Irvine,\\
	California 92697, USA\\
	E-mail: \email{mvalli@uci.edu}}

\abstract{Motivated by the persisting ``anomaly'' in the measurement of $P_5^\prime$, we review hadronic uncertainties entering the angular observables of the decay $\bar B\to {\bar K}^*\mu^+\mu^-$. We argue that hadronic uncertainties could account for the present measurements. We discuss how to extract information on the
non-factorizable hadronic contribution from experimental data exploiting its $q^2$ dependence and propose a parametrization optimized for this purpose. While no clear conclusion can be drawn with present experimental uncertainties, we show that future measurements should be able to pin down many hadronic parameters that we define in our parametrization.}

\FullConference{The International Conference on B-Physics at Frontier Machines - BEAUTY2018\\
		6-11 May, 2018\\
		La Biodola, Elba Island, Italy}

\begin{document}

\section{Introduction}
Anomalies in $B$ physics are attracting a lot of attention as of late. Deviations from the Standard Model (SM) predictions exceeding 3$\sigma$ persist in several observables, including exciting hints of lepton flavour universality (LFU) violation~\cite{Aaij:2015oid,Abdesselam:2016llu,Aaij:2014ora,Aaij:2017vbb}. Furthermore, the emerging pattern may have a simple explanation in terms of new physics (NP) contributions to the Wilson coefficients of one/few operator(s), which are obtained for instance in models with $Z^\prime$ or leptoquarks, for a recent review see the contributions of F.~Feruglio and A.~Greljo, or e.g. ref.~\cite{Marzocca:2018spk}.

In these proceedings, we focus on the prediction of the observable $P_5^\prime$ obtained from the angular analysis of the decay $\bar B\to {\bar K}^*\mu^+\mu^-$ in the low $q^2$ region, arguing that non-factorizable hadronic contributions coming from four-quark operators are not fully under control and could be responsible for the observed deviation from the SM. Elaborating on our previous results~\cite{Ciuchini:2015qxb,Ciuchini:2017mik}, we point out that the $q^2$ dependence can be used to extract from data interesting information on the hadronic contribution, even though it cannot be fully disentangled from the NP contribution in the absence of an adequate theoretical control. We then propose a variant of our previous parametrization of the hadronic contribution optimized for extracting information from data and show that, while no sound result can be obtained with present experimental uncertainties, yet future measurements should be able to pin down most hadronic parameters.

These proceedings are organized as follows. In section~\ref{sec:p5p} we set up the scene for the calculation of $P_5^\prime$. In section~\ref{sec:hlam} we focus on the non-factorizable hadronic contributions and present our new parametrization. The extraction of the hadronic parameters using present and (expected) future data are collected in section~\ref{sec:results}, while conclusions are drawn in section~\ref{sec:conclusions}.

\section{Calculating $P_5^\prime$}
\label{sec:p5p}
The angular analysis of the decay $\bar B\to {\bar K}^*\mu^+\mu^-$ allows to extract the coefficients $I_i$ of the fully differential decay rate
\begin{eqnarray}
\frac{d^4\Gamma}{dq^2\,d\!\cos{\theta_\ell}\,d\!\cos{\theta_K}\,d\phi} &\:\:=\:\:& \frac{9}{32\pi} \Big( I_1^s\sin^2{\theta_K} + I_1^c\cos^2{\theta_K} + (I_2^s\sin^2{\theta_K} + I_2^c\cos^2{\theta_K})\cos{2\theta_{\ell}} \nonumber \\
&& \hspace{-0.035\textwidth}+I_3\sin^2{\theta_K}\sin^2{\theta_{\ell}}\cos{2\phi} + I_4\sin{2\theta_K}\sin{2\theta_{\ell}}\cos{\phi}+ I_5\sin{2\theta_K}\sin{\theta_{\ell}}\cos{\phi}  \nonumber\\
&& \hspace{-0.035\textwidth}+ (I_6^s\sin^2{\theta_K} + I_6^c\cos^2{\theta_K})\cos{\theta_{\ell}}+ I_7\sin{2\theta_K}\sin{\theta_{\ell}}\sin{\phi}  \nonumber \\
&& \hspace{-0.035\textwidth}+ I_8\sin{2\theta_K}\sin{2\theta_{\ell}}\sin{\phi}+ I_9\sin^2{\theta_K}\sin^2{\theta_{\ell}}\sin{2\phi} \Big)\,.
\label{degamma}
\end{eqnarray}
These coefficients are conveniently recast in terms of the optimized variables $P_i$~\cite{Kruger:2005ep,Egede:2008uy,Descotes-Genon:2013vna}
written in terms of the CP-averaged angular coefficients $\Sigma_i = (I_i + \bar{I}_i)/2$. In particular, one defines
\begin{equation}
P_5^\prime = \frac{\Sigma_5}{2\sqrt{-\Sigma_{2s}\Sigma_{2c}}}\,.
\end{equation}

In the helicity basis~\cite{Melikhov:1998cd} adopted in these proceedings, the angular coefficients can be computed in terms of
seven helicity amplitudes~\cite{Jager:2012uw,Ciuchini:2015qxb}:
\begin{eqnarray}
H_V^{\lambda} &\:\:=\:\:& i \frac{4G_Fm_B}{\sqrt{2}} \frac{e^2}{16\pi^2}\lambda_t \left\{C^{\mathrm{eff}}_9\tilde{V}_{L\lambda}  + \frac{m_B^2}{q^2} \left[\frac{2m_b}{m_B}C_7^{\mathrm{eff}}\tilde{T}_{L\lambda}  - 16\pi^2h_{\lambda} \right]\right\},\nonumber\\
H_A^{\lambda} &=& i \frac{4G_Fm_B}{\sqrt{2}} \frac{e^2}{16\pi^2}\lambda_t C_{10}\tilde{V}_{L\lambda}, ~~~
H_P = -i \frac{4G_Fm_B}{\sqrt{2}} \frac{e^2}{16\pi^2}\lambda_t \frac{2m_{\mu}m_b}{q^2}  C_{10} \left( \tilde{S}_{L} - \frac{m_s}{m_B}\tilde{S}_{R} \right) \label{Hp},
\label{eq:helamp}
\end{eqnarray}
with $\lambda=0,\pm$. The CKM factor $\lambda_t=V_{ts}V_{tb}^*$, $C^\mathrm{eff}_{7,9,10}$ are Wilson coefficients of the $\Delta B=1$ effective weak Hamiltonian, $\tilde{V}_{L\lambda}$, $\tilde{T}_{L\lambda}$, $\tilde{S}_{L}$, $\tilde{S}_{R}$ are form factors entering the factorized part of the amplitudes (as defined in Appendix A of ref.~\cite{Ciuchini:2015qxb}), while $h_{\lambda}$ are the genuine non-factorizable hadronic contributions.

The detailed discussion of hadronic uncertainties related to form factors is beyond the scope of these proceedings, but we briefly comment on the current status. Although only a light-cone sum rules (LCSR) calculation of the form factors is available in the large recoil (low $q^2$) region~\cite{Straub:2015ica}, it matches reasonably well the extrapolation of lattice QCD calculations at low recoil~\cite{Horgan:2013hoa}. Moreover, LCSR results for the form factors are provided together with the full correlation matrix to allow taking into account correlations induced by the heavy quark symmetry. Thus the uncertainty of 10--15\% attached to the form factors looks credible and moreover it is further reduced in the optimized observables, making this contribution to the theoretical uncertainty quite smaller than the present experimental error.

In the next section we focus on the main topic of these proceedings, namely the non-factorizable contributions $h_{\lambda}$.

\section{Non-factorizable hadronic contributions}
\label{sec:hlam}
The non-factorizable hadronic contribution~\footnote{In the following we use the notation of ref.~\cite{Jager:2012uw}.}
\begin{equation}
 h_\lambda(q^2) = \frac{\epsilon^*_\mu(\lambda)}{m_B^2} \int d^4x e^{iqx} \langle \bar K^* \vert T\{j^{\mu}_\mathrm{em} (x) 
 \mathcal{H}_\mathrm{eff}^\mathrm{had} (0)\} \vert \bar B \rangle
 \label{eq:hlambda}
\end{equation}
is generated by the insertion in the matrix element of the four-quark operators present in the $\Delta B=1$ effective weak Hamiltonian, denoted
here as $\mathcal{H}_\mathrm{eff}^\mathrm{had}$, together with an electromagnetic quark current. Details on the definition of the effective Hamiltonian
can be found for instance in ref.~\cite{Ciuchini:2015qxb}.The largest contribution is given by the current-current
operators
\begin{equation}
  Q^c_1 = (\bar{s}_L\gamma_{\mu}T^a c_L)(\bar{c}_L\gamma^{\mu}T^ab_L)\,,\quad
  Q^c_2 = (\bar{s}_L\gamma_{\mu} c_L)(\bar{c}_L\gamma^{\mu}b_L)\,,
\end{equation}
with the two charm quark fields closed in a loop. These contributions are notoriously troublesome to estimate, as they can produce on-shell intermediate
hadronic states which give raise to strong phases, non-local amplitudes, etc. Many years ago, similar charm-loop contributions~\cite{Colangelo:1989gi,Ciuchini:1997hb} stimulated an intense debate about the validity of factorization in the infinite mass limit for heavy-to-light non-leptonic $B$ decays~\cite{Bauer:2004tj,Beneke:2004bn,Bauer:2005wb,Beneke:2009az}. For $B\to V \ell\ell$, factorization of the amplitudes in the infinite mass limit
has been proven in ref.~\cite{Beneke:2001at} at low $q^2$. Yet the issue of computing the non-factorizable contribution in eq.~(\ref{eq:hlambda}), albeit power suppressed, remains open.

The only estimate of $h_\lambda$ presently available can be found in ref.~\cite{Khodjamirian:2010vf}. Using LCSR, the authors of ref.~\cite{Khodjamirian:2010vf}
were able to compute $h_\lambda(q^2)$ for $q^2 \ll 4m_c^2$, where the single soft gluon approximation used in the calculation is applicable. These results
were then extended to all $q^2$ with a dispersion relation using a spectral function including the $J/\psi$ and $\psi^\prime$ resonances plus an additional
pole modeling the contribution from higher resonances and continuum. This combination of methods and approximations testifies the complexity of the calculation
which is reflected in the large uncertainty quoted by the authors, albeit intrinsic limitations of the adopted methods (e.g. lack of strong phases) and model dependence can hardly be quantified. The correction to $P_5^\prime$ induced by the result of ref.~\cite{Khodjamirian:2010vf} is not large, rather flat in $q^2$, and goes in the direction of increasing the anomaly.

Recently, attempts at confirming or improving the results of ref.~\cite{Khodjamirian:2010vf} have appeared in the literature~\cite{Bobeth:2017vxj,Blake:2017fyh}. The empirical model of ref.~\cite{Blake:2017fyh} assumes that $h_\lambda$ can be obtained as a sum of relativistic Breit-Wigner functions and uses resonance data to fix the parameters, although some of them, notably strong phases, cannot be fixed with present data. The result is remarkably in agreement with ref.~\cite{Khodjamirian:2010vf} for vanishing strong phases, but quite different for other choices, showing the importance of controlling strong phases. A more theoretical approach was followed in ref.~\cite{Bobeth:2017vxj}: the authors studied the analytic properties of $h_\lambda$, isolated the resonance poles and proposed a $z$ expansion for the remainder function, mapping the contribution of the cut at the boundary of the region of convergence. The coefficients of the expansion are fixed using both resonance data and LCSR results at negative $q^2$ provided by the authors of ref.~\cite{Khodjamirian:2010vf}. Results compare well with ref.~\cite{Khodjamirian:2010vf}, but the coefficients obtained at different orders show a poor convergence of the series.

Given what is at stake, we consider the present theoretical knowledge of $h_\lambda$ not fully satisfactory. LCSR estimates could be in the right ballpark, but the unsatisfying control over strong phases, the theoretical problems pointed out in ref.~\cite{Kozachuk:2018yxf}, and the fact that the $P_5$ anomaly lies in the $q^2$ region approaching the $J/\psi$ resonance call for extra care. We therefore decided to expand $h_\lambda$ in powers of $q^2$ in the region $q^2 \in [0,8]$ GeV$^2$ and use the $B\to K^*\mu\mu$ and $B\to K^*\gamma$ data to fix the coefficients, considering two cases: a ``standard'' scenario, denoted as $PMD$ (phenomenological model driven), where the results of ref.~\cite{Khodjamirian:2010vf} are used to constrain the coefficients of the expansion in the whole low $q^2$ region, and a ``conservative'' scenario, denoted as $PDD$ (phenomenological data driven), where only the actual LCSR results computed at $q^2=0$ and $1$ GeV$^2$ in ref.~\cite{Khodjamirian:2010vf} are used to constrain the absolute values of the $h_\lambda$, while their phases and $q^2$ dependence are inferred from the experimental data. In the second scenario, the constraining power
of $B\to K^*\mu\mu$ on NP is lost, as some coefficients of the $h_\lambda$ expansion are indistinguishable from NP contributions. However, one can still determine
most coefficients of $h_\lambda$ and look for ``unexpected'' hadronic contributions, to either invalidate or gain more confidence in the available estimates.

To this end, in these proceedings we propose a variation of the simple Taylor expansion of $h_\lambda$ we used in previous publications~\cite{Ciuchini:2015qxb,Ciuchini:2017mik} that reads~\footnote{The two definitions of each $h_\lambda$ are equivalent
up to higher order terms in the $q^2$ expansion. Notice the different $q^2$ behaviour of $h_0$~\cite{Arbey:2018ics}.}
\begin{eqnarray}
h_-(q^2) &=& -\frac{m_b}{8\pi^2 m_B} \tilde T_{L -}(q^2) h_-^{(0)} -\frac{1}{16\pi^2 m_B^2} \tilde V_{L -}(q^2) h_-^{(1)} q^2 + h_-^{(2)} q^4+{\cal O}(q^6)\nonumber\\
&=&-\frac{1}{16\pi^2}\left[\frac{2m_b}{m_B} \tilde T_{L -}(0) h_-^{(0)} +\left(\frac{1}{m_B^2} \tilde V_{L -}(0) h_-^{(1)} + \frac{2 m_b}{m_B} \frac{d\tilde T_{L -}}{dq^2}(0) h_-^{(0)}\right) q^2\right.\nonumber\\
& & \qquad\qquad\left.+\left(\frac{m_b}{m_B} \frac{d^2\tilde T_{L -}}{(dq^2)^2}(0) h_-^{(0)} + \frac{1}{m_B^2}\frac{d\tilde V_{L -}}{dq^2}(0) h_-^{(1)} -16\pi^2 h_-^{(2)}\right) q^4\right]+{\cal O}(q^6)\,,\nonumber\\
h_+(q^2) &=&  h_+^{(0)}-\frac{m_b}{8\pi^2 m_B} \tilde T_{L +}(q^2) h_-^{(0)} +\left(h_+^{(1)}-\frac{\tilde V_{L +}(q^2)}{16\pi^2 m_B^2}  h_-^{(1)}\right) q^2 + h_+^{(2)} q^4+{\cal O}(q^6)\nonumber \\
&=& -\frac{1}{16\pi^2}\left[\frac{2m_b}{m_B} \tilde T_{L +}(0) h_-^{(0)} - 16\pi^2 h_+^{(0)} +\left(\frac{1}{m_B^2} \tilde V_{L +}(0) h_-^{(1)} + \frac{2m_b}{m_B} \frac{d\tilde T_{L +}}{dq^2}(0) h_-^{(0)}\right.\right.\nonumber\\
& & \quad\left.\left.- 16\pi^2 h_+^{(1)}\right) q^2+\left(\frac{m_b}{m_B} \frac{d^2\tilde T_{L +}}{(dq^2)^2}(0) h_-^{(0)} + \frac{1}{m_B^2}\frac{d\tilde V_{L +}}{dq^2}(0) h_-^{(1)} - 16\pi^2 h_+^{(2)}\right) q^4\right]+{\cal O}(q^6)\,,\nonumber \\
h_0(q^2) &=& h_0^{(0)}\sqrt{q^2}-\frac{m_b}{8\pi^2 m_B} \tilde T_{L 0}(q^2) h_-^{(0)} -\frac{\tilde V_{L 0}(q^2)}{16\pi^2 m_B^2}  h_-^{(1)} q^2 +h_0^{(1)}(q^2)^\frac{3}{2}  +{\cal O}((q^2)^\frac{5}{2})\nonumber \\
&=& -\frac{1}{16\pi^2}\left[\left(\frac{2m_b}{m_B} \lim_{q^2\to 0}\left(\frac{\tilde T_{L 0}}{\sqrt{q^2}}\right) h_-^{(0)}+\frac{1}{m_B^2}\lim_{q^2\to 0}{\left(\sqrt{q^2}\tilde V_{L 0}\right)}h_-^{(1)} - 16\pi^2 h_0^{(0)}\right) \sqrt{q^2}~+\right.\nonumber\\
& & \!\left.\left(\frac{2 m_b}{m_B}\frac{d}{dq^2}\frac{\tilde T_{L 0}}{\sqrt{q^2}}(0) h_-^{(0)}+\frac{1}{m_B^2}\frac{d\sqrt{q^2}\tilde V_{L 0}}{dq^2}(0)h_-^{(1)}-16\pi^2  h_0^{(1)}\right) (q^2)^{\frac{3}{2}}\right]+{\cal O}\left((q^2)^{\frac{5}{2}}\right)\,,
\label{eq:newexp}
\end{eqnarray}
such that the contributions to the helicity amplitudes $H_V^\lambda$ become
\begin{eqnarray}
H_V^{-} &\:\:\propto\:\:& \left\{\left(C^{\mathrm{eff}}_9 + 
h_-^{(1)}\right)\tilde V_{L -}  + \frac{m_B^2}{q^2} 
\left[\frac{2m_b}{m_B}\left(C_7^{\mathrm{eff}} + h_-^{(0)} \right) 
\tilde T_{L -} - 16\pi^2 h_-^{(2)}\, q^4 \right]\right\}\,, \nonumber\\
H_V^{+} &\:\:\propto\:\:&  \left\{\left(C^{\mathrm{eff}}_9 + 
h_-^{(1)}\right)\tilde V_{L +}  + \frac{m_B^2}{q^2} 
\left[\frac{2m_b}{m_B}\left(C_7^{\mathrm{eff}} + h_-^{(0)} \right) \tilde T_{L +}
-  16\pi^2\left(h_{+}^{(0)} + h_{+}^{(1)}\, q^2 +  h_{+}^{(2)}\, q^4\right) \right]\right\}\,,\nonumber\\
H_V^{0} &\:\:\propto\:\:&  \left\{\left(C^{\mathrm{eff}}_9 + h_-^{(1)}\right) \tilde{V}_{L0}  + 
\frac{m_B^2}{q^2} \left[\frac{2m_b}{m_B} \left(C_7^{\mathrm{eff}} + h_-^{(0)} \right) 
\tilde{T}_{L0}-16\pi^2\sqrt{q^2}\left({h}_{0}^{(0)} + {h}_{0}^{(1)}\, q^2\right)\right]\right\}\,.
\label{eq:hv}
\end{eqnarray}
The equations above clearly show that $h_-^{(0)}$ and $h_-^{(1)}$ are constant shifts to the the Wilson coefficients $C^{\mathrm{eff}}_{7,9}$ that cannot be distinguished from NP contributions. Therefore, one cannot fit $h_-^{(0)}$ and $h_-^{(1)}$ from data without assuming the validity of the SM and conversely one cannot establish NP from data without a theory input on these coefficients. On the other hand, all the other coefficients $h_\lambda^{(i)}$ can in principle be fitted.

In the next section, we will present the determination of the coefficients $h_\lambda^{(i)}$ from present experimental data. We discuss the perspective of this analysis with improved data and comment on the impact of our approach on the NP interpretation of the $B$ anomalies.

\begin{figure}[t]
  \centering
  \includegraphics[width=.43\textwidth]{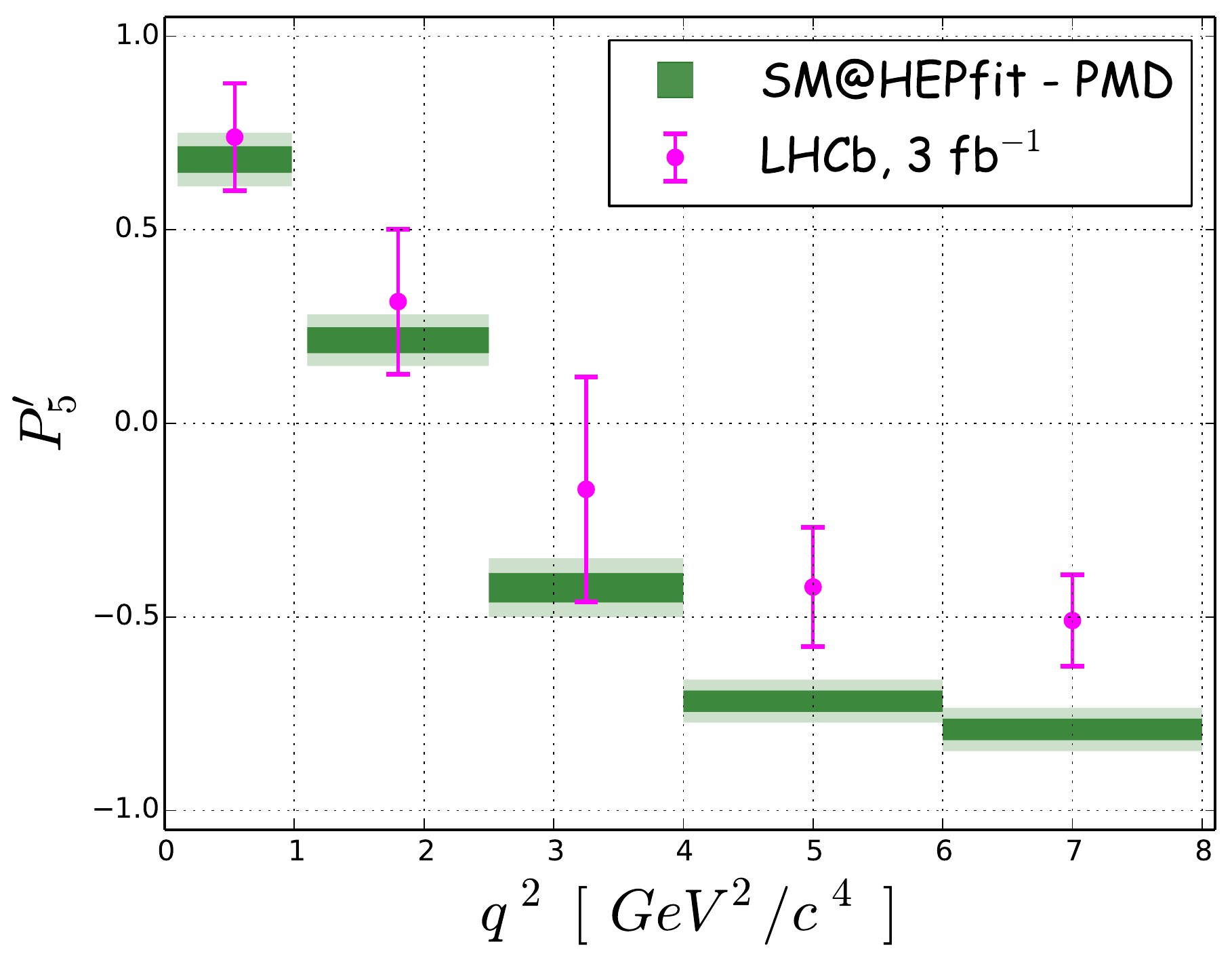}
  \hspace{0.04\textwidth}
  \includegraphics[width=.43\textwidth]{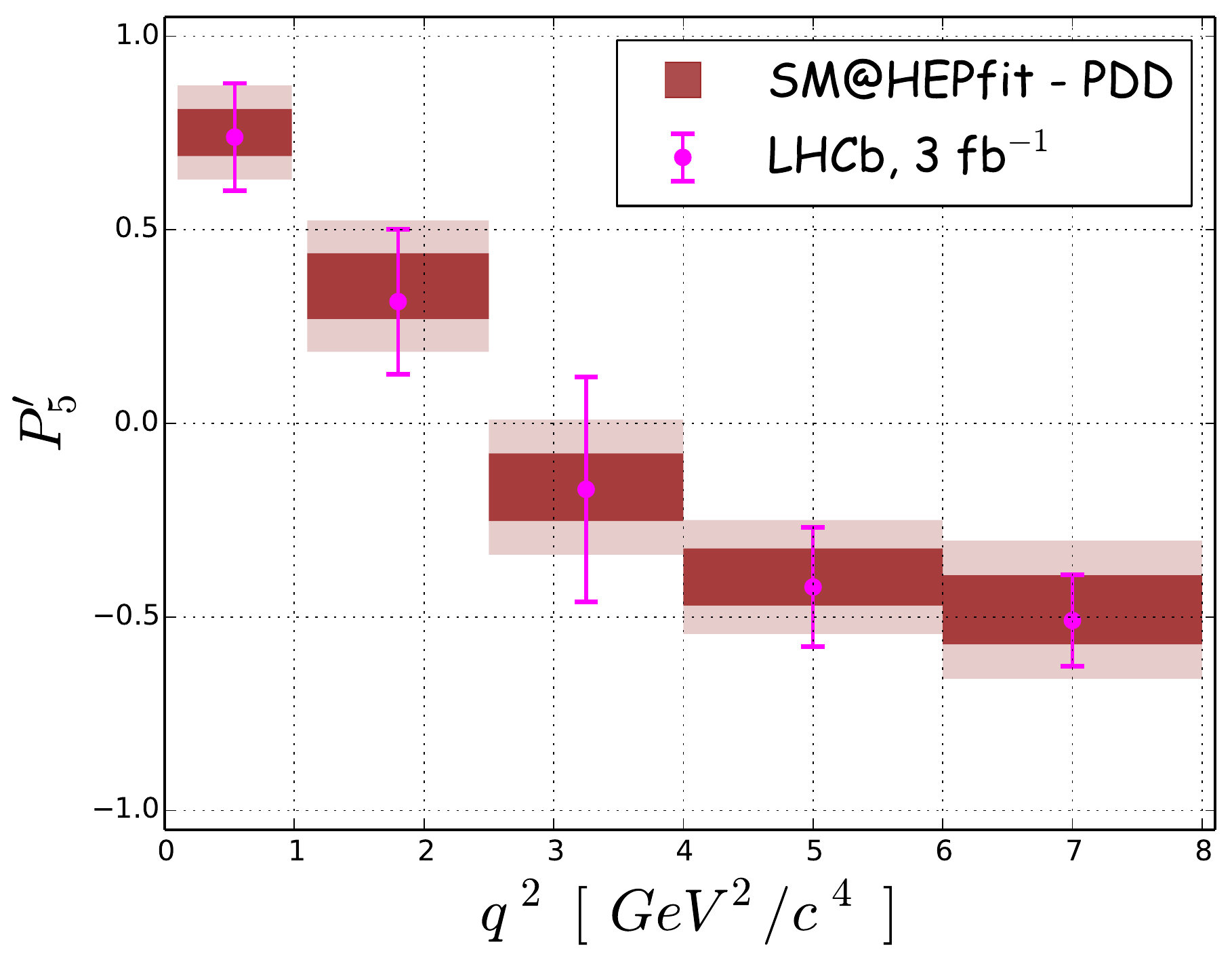}
  \caption{The posterior of $P_5^\prime$ in the $PMD$ (left panel) and $PDD$ (right panel) fits compared with the LHCb measurements.}
  \label{fig:p5p}
\end{figure}

\section{Present fit and extrapolations}
\label{sec:results}
We present results from a global analysis of the $B$ anomalies, along the lines of the one presented in ref.~\cite{Ciuchini:2017mik}, but we focus here
on $P_5^\prime$ and the hadronic parameters $h_\lambda^{(i)}$. Inputs updated since ref.~\cite{Ciuchini:2017mik} are listed in table~\ref{tab:tab}.
Let us first comment on the determination of $P_5^\prime$ from the SM analysis in the two considered cases: as shown in the left panel of fig.~\ref{fig:p5p},
the anomaly is clearly present in the $PMD$ case, where the results of ref.~\cite{Khodjamirian:2010vf} are used to constrain the coefficients of
the expansion in eq.~(\ref{eq:newexp}) over all the considered $q^2$ range, while it is no longer present in the $PDD$ fit (right panel of fig.~\ref{fig:p5p})
where the $q^2$ dependence of $h_\lambda$ is unconstrained and determined from data. As shown in refs.~\cite{Ciuchini:2016weo,Ciuchini:2017gva}, the fitted correction is large in the $q^2$ bins where the anomaly is, but still compatible with a power suppressed correction. We then conclude that the evidence for the $P_5^\prime$ anomaly is fully based on the only available estimate of $h_\lambda$ in ref.~\cite{Khodjamirian:2010vf}, with the caveats we discussed above.

\begin{table}[t]
\centering
\begin{tabular}{c|c|c|c}
\hline
$f_{K^*,\parallel}$ [MeV] & 
$f_{K^*,\perp}$ [MeV] & 
$a_1(\bar{K}^*)_{\perp,\parallel}$ &
$a_2(\bar{K}^*)_{\perp,\parallel}$ \\
\hline
$204 \pm 7$ & $159 \pm 6$ & $0.04 \pm 0.03$ & $0.11 \pm 0.09$ \\
\hline
\end{tabular}
\caption{Inputs of our global analysis which has been updated with
respect to ref.~\cite{Ciuchini:2017mik}.}
\label{tab:tab}
\end{table}

\begin{figure}[t]
  \centering
  \includegraphics[width=.47\textwidth]{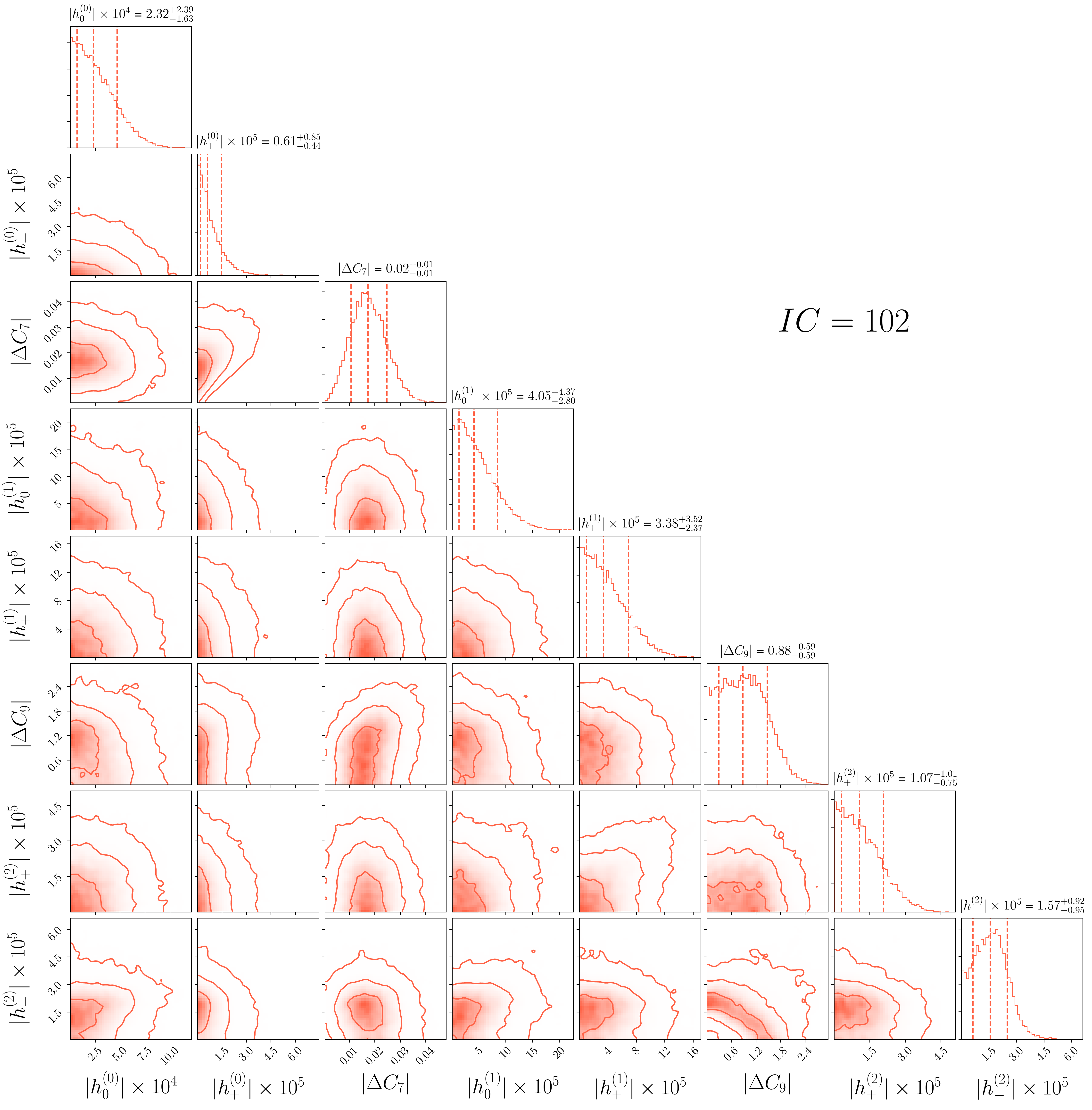}
  \hspace{0.04\textwidth}
  \includegraphics[width=.47\textwidth]{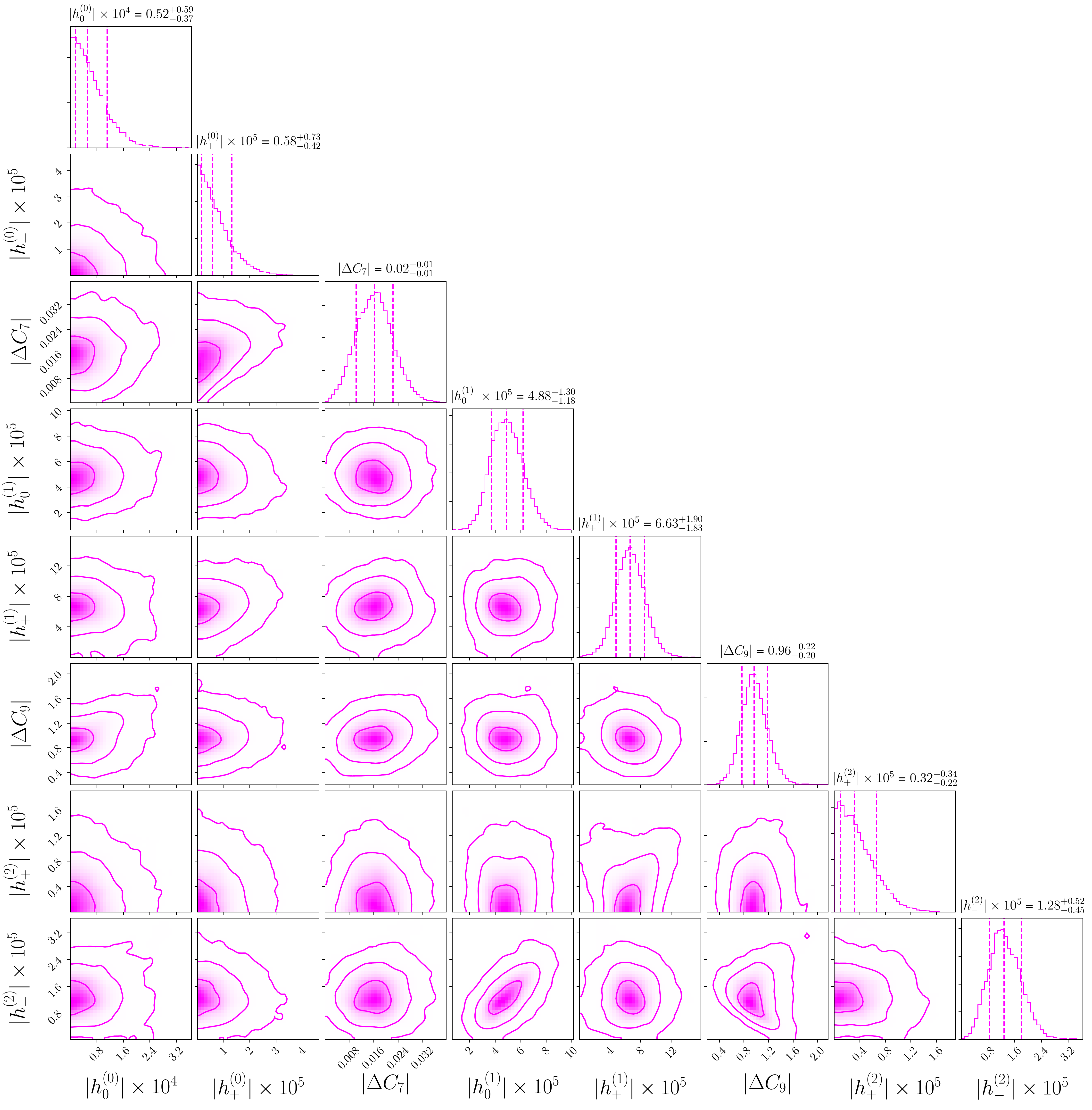}
  \caption{Determination of the coefficients $|h_\lambda^{(i)}|$ from present (left panel) and future (right panel) data. See text for details.}
  \label{fig:hlamcoeff}
\end{figure}

\begin{figure}[t]
  \centering
  \includegraphics[width=.4\textwidth]{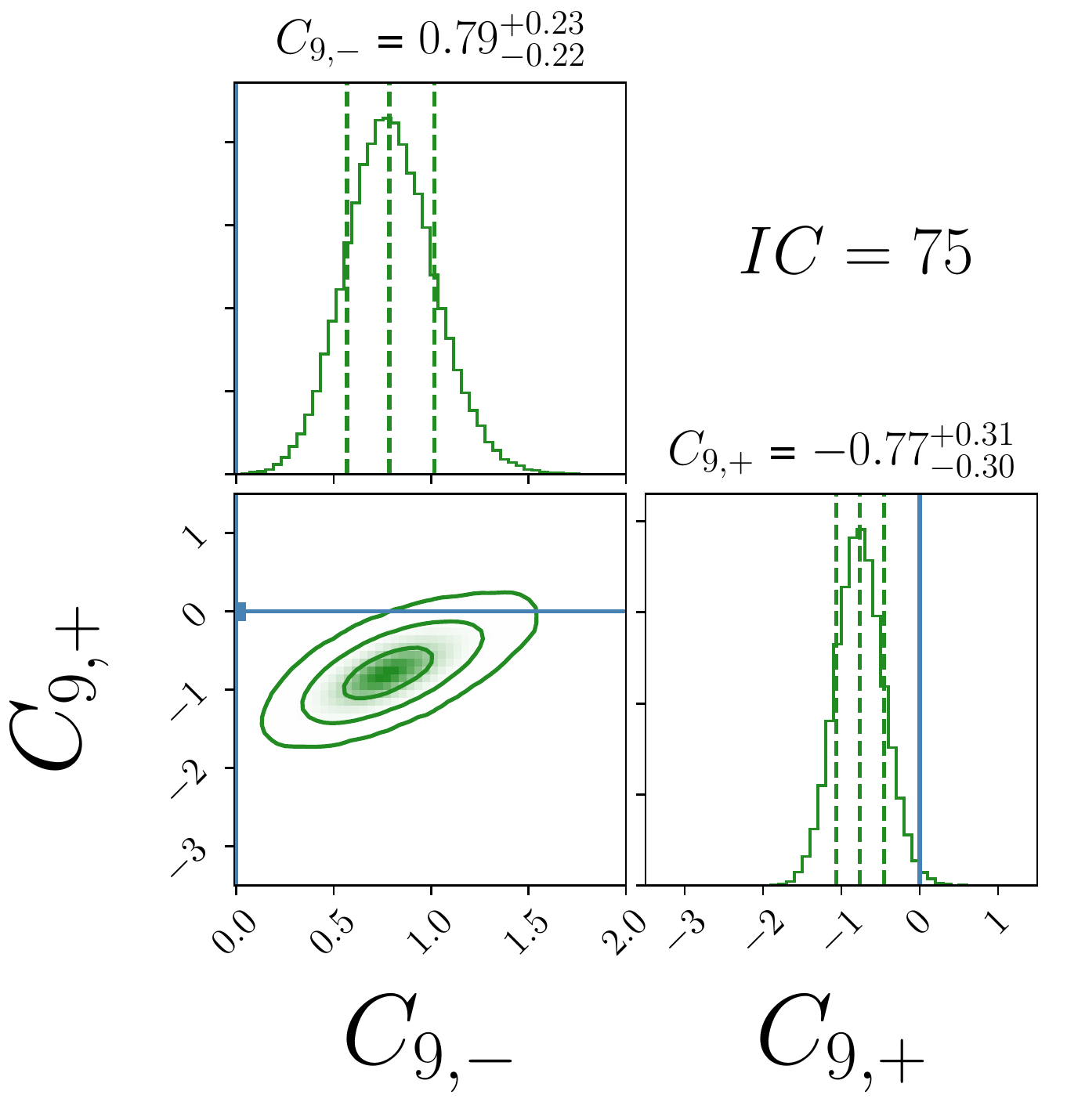}
  \hspace{0.04\textwidth}
  \includegraphics[width=.4\textwidth]{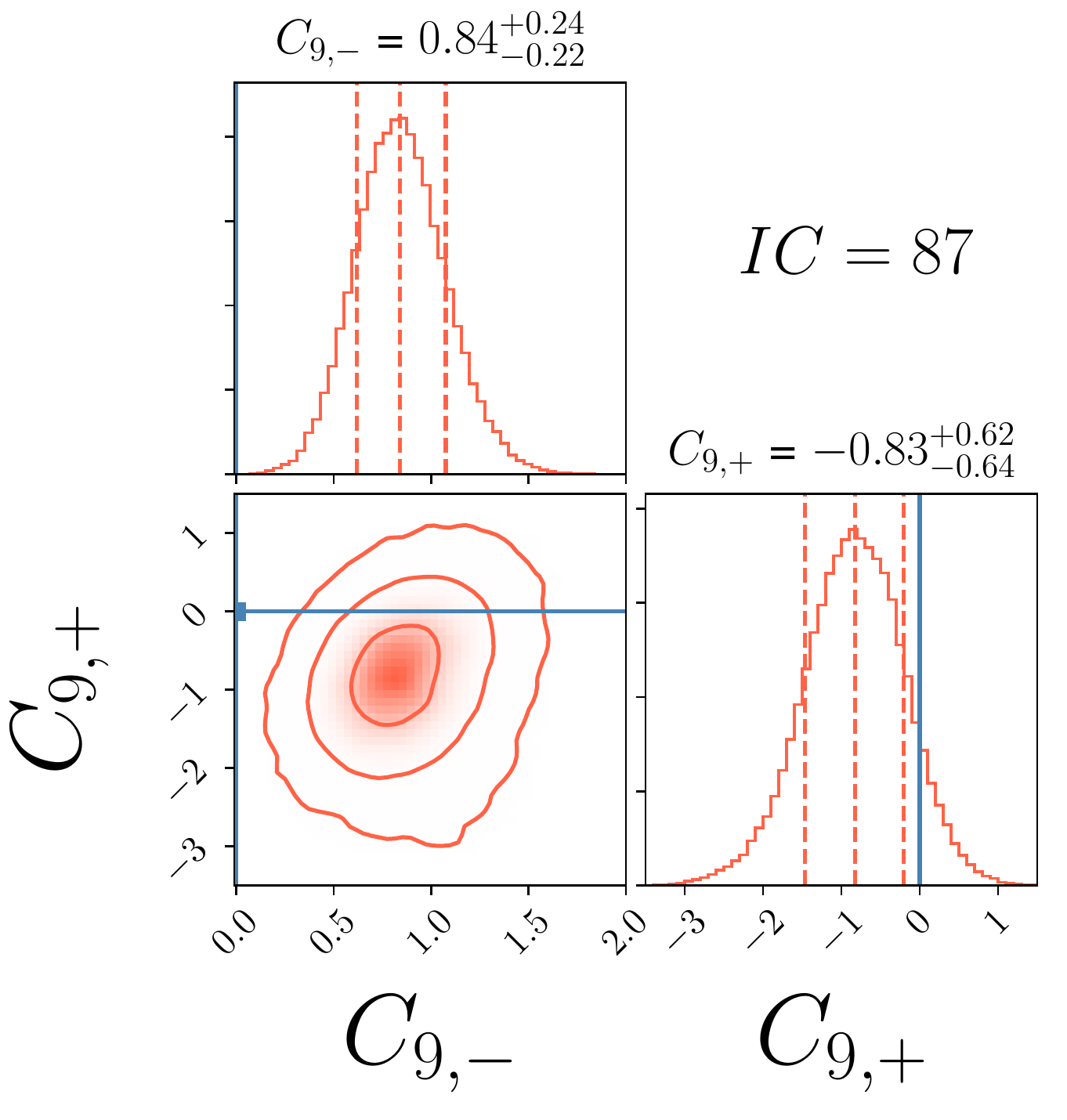}
  \includegraphics[width=.4\textwidth]{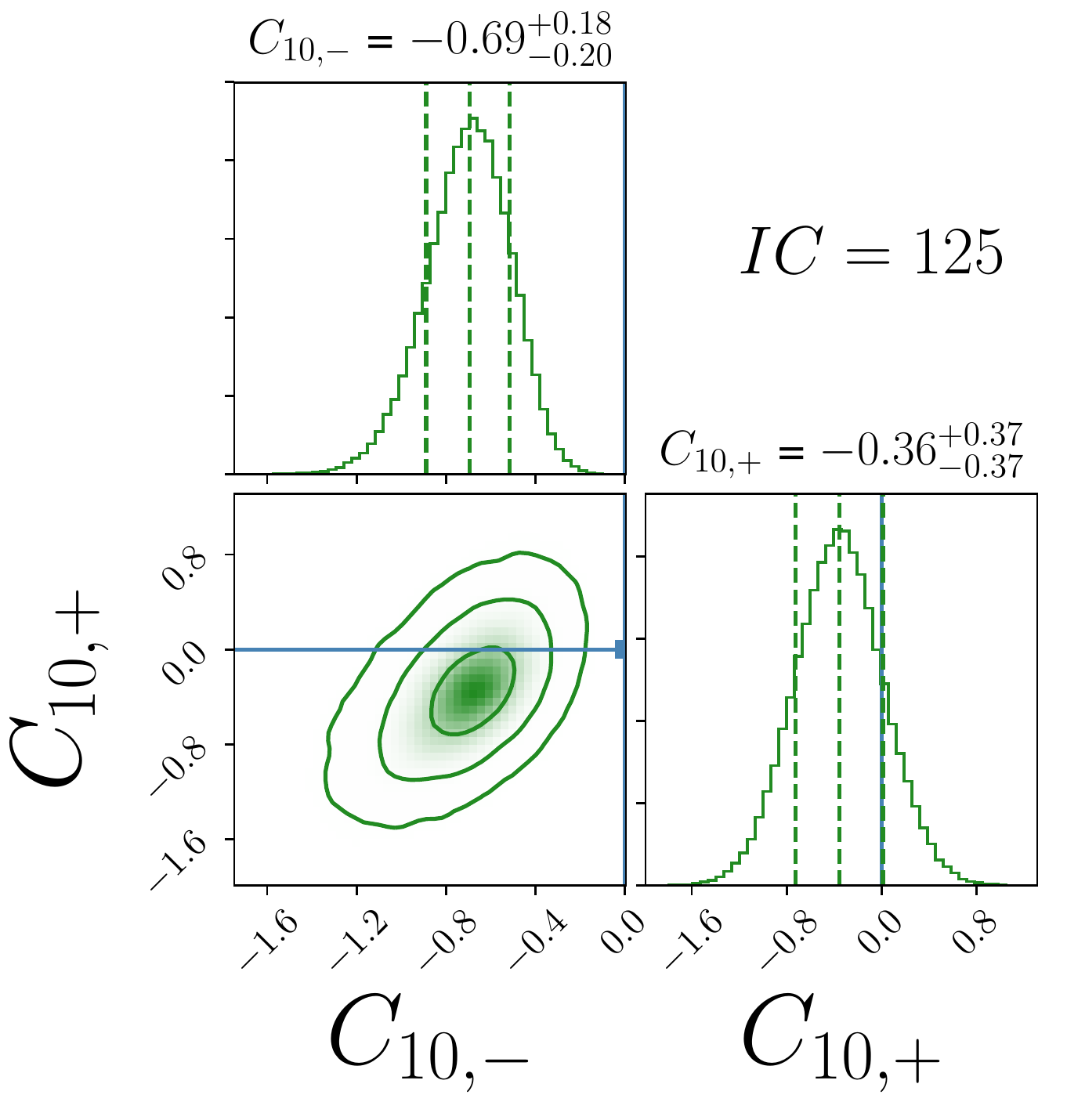}
  \hspace{0.04\textwidth}
  \includegraphics[width=.4\textwidth]{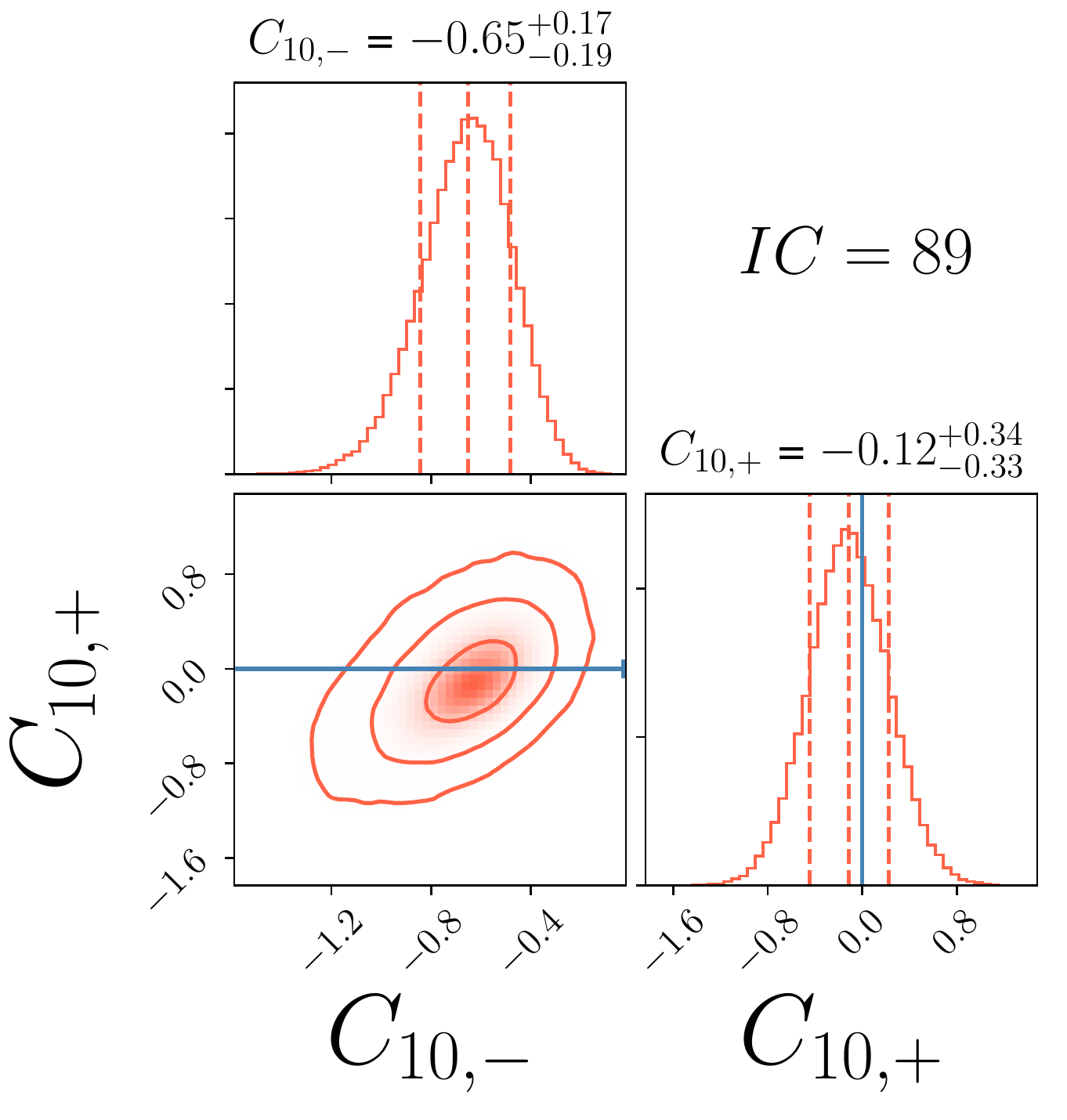}
  \caption{Upper row: global fit for $C^\mathrm{NP}_{9,\pm}$ in the $PMD$ (left) and $PDD$ (right) case. Lower row: global fit for $C^\mathrm{NP}_{10,\pm}$ in the $PMD$ (left) and $PDD$ (right) case.}
  \label{fig:c+c-}
\end{figure}

Let us now move on to the determination of the coefficients of the $h_\lambda$ expansion. In the left panel of fig.~\ref{fig:hlamcoeff} we show the determination
of absolute values of the coefficients $h_\lambda^{(i)}$ together with the correlations from the SM fit to present data in the $PDD$ case. The fit is not good (as signaled by the value of the information criterion (IC) in fig.~\ref{fig:hlamcoeff} compared to the NP fits in fig,~\ref{fig:c+c-}, keeping in mind that smaller values correspond to better fits), as LFU-violating anomalies cannot be accommodated in the SM. However, as we have discussed above, the posterior of $P_5^\prime$ agrees with the measurement. From the plot, we can conclude that the present experimental uncertainties do not allow a clear determination of the hadronic parameters (a similar conclusion holds for the phases). Indeed the only parameter clearly different from zero is $|h_-^{(0)}|$ (denoted as $|\Delta C_7|$ in the plot), as a consequence of imposing the constraint from the theoretical estimate of ref.~\cite{Khodjamirian:2010vf} at $q^2=0$. There is however an interesting correlation between $|h_-^{(1)}|=|\Delta C_9|$ and $|h_-^{(2)}|$: the present anomaly can be reproduced either with a constant shift of the
Wilson coefficient $C^\mathrm{eff}_9$ (due to hadronic contribution or NP, no way to disentangle them) or with a $q^4$ term in $h_-$. If $|\Delta C_9|$ is small,
$|h_-^{(2)}|$ is found to be different from zero at more than 2$\sigma$, in agreement with the finding of ref.~\cite{Ciuchini:2015qxb}.

We then repeated the analysis by reducing the experimental error by a factor of six, with the central values given by the global mode of the SM fit. The errors obtained in this simple way are in the ballpark of what is expected from the future LHCb upgrade. From the right panel of fig.~\ref{fig:hlamcoeff}, it can be seen that many coefficients $|h_-^{(i)}|$
can be extracted from data once the experimental error is reduced. In particular, the correlation between $|\Delta C_9|$ and $|h_-^{(2)}|$ is much reduced, allowing to distinguish a constant shift of the Wilson coefficient from a rise of $h_\lambda$ for $q^2=6$--$8$ GeV$^2$. A more detailed analysis, including a discussion of the phases, will be presented in a forthcoming publication.

Before concluding, we comment on the effect of the non-factorizable hadronic contributions on the NP interpretation of the $B$ anomalies. In fig.~\ref{fig:c+c-}, we
plot the NP coefficients
\begin{equation}
    C_{9,\pm}^\mathrm{NP} =\frac{1}{2} \left(C_{9,\mu}^\mathrm{NP}\pm C_{9,e}^\mathrm{NP}\right)\,,\quad
    C_{10,\pm}^\mathrm{NP} =\frac{1}{2} \left(C_{10,\mu}^\mathrm{NP}\pm C_{10,e}^\mathrm{NP}\right)\,,
\end{equation}
as fitted in the $PMD$ and $PDD$ cases. It is shown that the LFU-conserving coefficients $C_{9, +}^\mathrm{NP}$ and $C_{10, +}^\mathrm{NP}$ are affected by the different treatment of the charm-loop contribution, while the LFU-violating coefficients $C_{9, -}^\mathrm{NP}$ and $C_{10, -}^\mathrm{NP}$ are not, as expected. For $C_{9,\pm}^\mathrm{NP}$, both $PMD$ and $PDD$ cases provide a good fit (the IC value of $PMD$ is smaller, reflecting the more economical description of the anomalies in terms of NP contributions only), but the evidence for a deviation of $C_{9,+}^\mathrm{NP}$ from zero is much less significant in the $PDD$ case, as the measurement of $P_5^\prime$ is accommodated by hadronic contributions. As for the explanation of the $B$ anomalies in terms of $C_{10,\pm}^\mathrm{NP}$, the effect of the charm-loop contribution is more striking: this scenario produces a bad fit in the $PMD$ case, as $C_{10}$ alone cannot account for the $P_5^\prime$ anomaly, but is perfectly viable in the $PDD$ case, where NP is not needed to reproduce $P_5^\prime$ (see IC's in fig.~\ref{fig:c+c-}), as pointed out in ref.~\cite{Ciuchini:2017mik}.

\section{Conclusions}
\label{sec:conclusions}
We have reviewed hadronic uncertainties entering the angular observables of the decay \linebreak ${\bar B\to {\bar K}^*\mu^+\mu^-}$, arguing that the non-factorizable hadronic contribution could account for the present measurements. We have proposed a new parametrization of this contribution optimized to fit the new parameters from data, exploiting the $q^2$ dependence of the correction. While a fit to present data produces no clear determination of many of these parameters, we have shown how future measurements could be able to pin down many of them, improving our knowledge of the theoretically challenging charm-loop contribution.
Finally, we have emphasized once more that the NP interpretation of the $B$ anomalies is affected by hadronic uncertainties, showing how an explanation in terms of
$C_{10}$ becomes viable if the charm-loop contribution is treated as we have suggested.

\section*{Acknowledgements}
The work of M.C. was performed in part at the Aspen Center for Physics, which is supported by National Science Foundation grant PHY-1607611, and was partially supported by a grant from the Simons Foundation. M.C. is associated to the Dip. di Matematica e Fisica, Universit\`a di Roma Tre, and E.F. and L.S. are associated to the Dip. di Fisica, Universit\`a di Roma “La Sapienza”.

\end{document}